\documentclass[runningheads,a4paper]{llncs}

\usepackage{amsfonts}
\usepackage{dsfont}

\usepackage{mathtools}

\usepackage{amssymb}

\usepackage{amsmath}

\usepackage{amsthm}

\usepackage{graphicx}

\usepackage{epstopdf}

\usepackage{enumitem} 

\PassOptionsToPackage{hyphens}{url}\usepackage{hyperref}

\usepackage{caption}

\usepackage{float}

\usepackage{subcaption}

\usepackage{epsfig}
\usepackage{mdframed}

\usepackage{multirow}

\usepackage{url}

\PassOptionsToPackage{hyphens}{url}\usepackage{hyperref}

\urldef{\mailsa}\path|{adnan.rashid, osman.hasan}@seecs.nust.edu.pk|

\newcommand{\keywords}[1]{\par\addvspace\baselineskip
\noindent\keywordname\enspace\ignorespaces#1}

\begin{document}

\mainmatter  

\title{Formal Analysis of Robotic Cell Injection Systems using Theorem Proving}

\titlerunning{Formal Analysis of Robotic Cell Injection Systems}

\author{Adnan Rashid \and Osman Hasan}

%
\authorrunning{A. Rashid and O. Hasan}


\institute{School of Electrical Engineering and Computer Science (SEECS)\\
National University of Sciences and Technology (NUST)\\
Islamabad, Pakistan\\
\mailsa\\
}

%
%
\maketitle

\begin{abstract}

Cell injection is an approach used for the delivery of small sample substances into a biological cell and is widely used in drug development, gene injection, intracytoplasmic sperm injection (ICSI) and in-virto fertilization (IVF). Robotic cell injection systems provide the automation of the process as opposed to the manual and semi-automated cell injection systems, which require expert operators and involve time consuming processes and also have lower success rates. The automation of the cell injection process is achieved by controlling the injection force and planning the motion of the injection pipette. Traditionally, these systems are analyzed using paper-and-pencil proof and computer simulation methods. However, the former is human-error prone and the later is based on the numerical algorithms, where the approximation of the mathematical expressions introduces inaccuracies in the analysis. Formal methods can overcome these limitations and thus provide an accurate analysis of the cell injection systems. Model checking, i.e., a state-based formal method, has been recently proposed for the analysis of these systems. However, it involves the discretization of the differential equations that are used for modeling the dynamics of the system and thus compromises on the completeness of the analysis of these safety-critical systems. In this paper, we propose to use higher-order-logic theorem proving, a deductive-reasoning based formal method, for the modeling and analysis of the dynamical behaviour of the robotic cell injection systems. The proposed analysis, based on the HOL Light theorem prover, enabled us to identify some discrepancies in the simulation and model checking based analysis of the same robotic cell injection system.

\keywords{Robotic Cell Injection System, Higher-order Logic, Theorem Proving}
\end{abstract}

\section{Introduction}\label{SEC:Intro}
Biological cell injection is a method used for the insertion of small amount of substances, i.e., bio-molecules, sperms, genes and proteins, into the suspended or adherent cells. It is widely used in gene injection~\cite{kuncova2004challenges}, drug development~\cite{nakayama1998new}, intracytoplasmic sperm injection (ISCI)~\cite{yanagida1999usefulness} and in-vitro fertilization (IVF)~\cite{sun2002biological}. For example, in IVF, the sperm is injected into matured eggs for the treatment of infertility. Similarly, drug development involves the injection of drugs into a cell and the observation of its implication at the cellular level.

Robotic cell injection systems can automatically perform the task of cell injection as opposed to the traditionally adopted manual and semi-automated injection procedures, which require trained operators and time-consuming processes and also have lower success rates. The most important factor in a robotic cell injection system is the injection force~\cite{huang2009visual} as a slight excessive force may damage the membrane of the cell~\cite{huang2006visual} or an insufficient force may not be able to pierce the cell~\cite{faroque2016virtual}.
Moreover, these robotic systems consist of an injection manipulator, digital cameras, sensors and microscope optics~\cite{huang2009visual} and thus the accuracy of the orientation and movement of these fundamental components is vital for the reliability of the overall system.
Thus, the robotic cell injection system designs need to be analyzed and verified quite carefully to ensure that these requirements are exhibited by the final systems.

The first step in the analysis of a robotic cell injection system is to model the coordinate frames corresponding to the orientations of its various components, i.e, the injection manipulator, cameras and images. This model allows us to capture the movement and thus the positions of these components during the process of cell injection. Moreover, the relationship between these coordinates provides the relative positions of these components, which is quite vital for a successful cell injection procedure. Next, in order to perform the process of injection, the motion planning of the injection pipette is modeled using some force control algorithms, such as the contact-space-impedance force control~\cite{sun1997modeling,huang2009visual} and the image-based torque controller~\cite{huang2006visual}. These controllers capture the overall dynamics of the system and are mainly responsible for the smooth functionality of the system during the process of cell injection.

Traditionally, the robotic cell injection systems have been analyzed using paper-and-pencil techniques. However, these manual analysis methods are prone to human error and also are not scalable for analyzing complex models like the robotic cell injection systems. Moreover, in some cases, all the required assumptions are not documented in the mathematical analysis, which may lead to erroneous design and analysis. Similarly, the computer simulations and the numerical methods have been used for the analysis of these systems. However, due to the continuous nature of the analysis and the limited amount of computer memory and the computational resources, the system is analyzed for a certain number of test cases only and thus the absolute accuracy cannot be achieved. Computer algebra systems, such as Mathematica~\cite{Mathematica2016webref}, have also been used for analyzing these systems~\cite{nethery1994robotica}. However, the symbolic algorithms residing in the core of these systems are unverified~\cite{duran2013misfortunes}, which puts a question mark on the accuracy of these analyses. Due to the safety-critical nature of robotic cell injection systems, the above-mentioned traditional techniques cannot be relied upon as they are either error prone or incomplete, which may lead to an undetected error in the analysis that may in turn lead to disastrous consequences.

Formal methods~\cite{hasan2015formal} are computer-based mathematical analysis techniques that can overcome the above-mentioned inaccuracies. Primarily, these techniques involve the development of a mathematical model of a system and verification of its properties using computer-based mathematical reasoning.
Sardar et al.~\cite{sardar2017towards} recently used probabilistic modeling checking~\cite{clarke1999model}, i.e., a state-based formal method, to formally analyze the robotic cell injection systems. However, their methodology involves the discretization of the differential equations that model the dynamics of these systems, which compromises the accuracy of the corresponding analysis. Moreover, the analysis also suffers from the inherent state-space explosion problem~\cite{clarke2012model}. Higher-order-logic theorem proving~\cite{harrison2009handbook} is an interactive verification technique  that can overcome these limitations. It primarily involves the mathematical modeling of the system based on higher-order logic and verification of its properties based on deductive reasoning. Given the high expressiveness of higher-order logic, it can truly capture the behavior of the differential equations, which is not possible in model checking based analysis.

In this paper, we propose to use the higher-order-logic theorem proving to formally analyze the robotic cell injection systems~\cite{huang2006visual} using the HOL Light theorem prover~\cite{harrison1996hol}.
The main motivation for the selection of HOL Light is the availability of reasoning support for real calculus~\cite{hol_light2017realanalysis}, multivariate calculus~\cite{hol_light2017multivariate}, vectors~\cite{hol_light2017vectors} and matrices~\cite{hol_light2017vectors}, which are some of the foremost requirements for formally analyzing robotic cell injection systems.
The major contributions of the paper are:

 \begin{itemize}[wide=0.5\parindent]
 \item[$\bullet$] \emph{Formalization of the cell injection system}, which includes the formal modeling of camera, stage and image coordinates and formal verification of their interrelationships in higher-order logic. It also includes the formal modeling of their dynamical behaviour (dynamics of two degrees of freedom (DOF) motion stage) using a system of differential equations and the formal verification of their solutions.
 \item[$\bullet$]  \emph{Formalization of the motion planning of the injection pipette}, which includes the formal modeling of the contact-space-impedance force control and the image-based torque controller and formal verification of their interrelationship.
 \item[$\bullet$] \emph{Identification of the discrepancies in the simulation and model checking based analysis of these systems}, i.e., the mathematical expression representing the image-based torque controller used in both simulation and model checking based analysis of the same system was found to be wrong based on the reported formalization in this paper.
 \end{itemize}

The rest of the paper is organized as follows: Section~\ref{SEC:prelim} provides an introduction about the HOL Light theorem prover, multivariate calculus theories of HOL Light and the robotic cell injection system. Section~\ref{SEC:form_analy_cell_inj_sys} presents the formalization of robotic cell injection system. We present the formalization of motion planning of the injection pipette in Section~\ref{SEC:form_mot_pla_inj_pip}. This also includes the identification of the discrepancies in the simulation and model checking based analysis of the same system. Finally, Section~\ref{SEC:conclusion} concludes the paper.

\section{Preliminaries} \label{SEC:prelim}

This section presents an introduction to the HOL Light theorem prover, multivariate calculus theories of HOL Light and the robotic cell injection system.

\subsection{HOL Light Theorem Prover} \label{SUBSEC:hol_light}

HOL Light~\cite{harrison1996hol} is a theorem proving environment that belongs to the family of HOL theorem provers. It is implemented in the meta language (ML)~\cite{paulson1996ml}, which is a functional programming language and is widely used for the construction of the mathematical proofs in the form of theories. A theory in HOL Light consists of types, constants, definitions, axioms and theorems. The HOL Light theories are ordered in a hierarchical fashion and the child theories can inherit the types, definitions and theorems of the parent theories. Every new theorem has to be verified based on the primitive inference rules and basic axioms or already verified theorems present in HOL Light, which ensures the soundness of this technique. HOL Light provides an extensive support for the analysis based on Boolean algebra~\cite{hol_light2017boolalgebra}, real arithmetics~\cite{hol_light2017realarith}, multivariable calculus~\cite{harrison2013hol} and vectors~\cite{hol_light2017vectors}. There are many automatic proof procedures~\cite{harrison1996formalized}, available in HOL Light, which are very useful in verifying the mathematical results automatically.


\subsection{Multivariable Calculus Theories in HOL Light} \label{SUBSEC:Mult_cal_theories}

A $\mathds{N}$-dimensional vector is represented as a $\mathds{R^N}$ column matrix with each of its element as a real number in HOL Light~\cite{harrison2013hol}. All of the vector operations are thus performed using matrix manipulations. Similarly, all of the multivariable calculus theorems are verified in HOL Light for functions with an arbitrary data-type $\mathds{R^N} \rightarrow \mathds{R^M}$.

Some of the frequently used HOL Light functions in the reported formalization are explained below:

\begin{mdframed}
\begin{definition}
\label{DEF:cx_and_ii}
\emph{Vector} \\{
\textup{\texttt{$\vdash$ $\forall$ l. vector l = (lambda i. EL (i - 1) l)
}}}
\end{definition}
\end{mdframed}

\noindent The function $\mathtt{vector}$ accepts a list \texttt{l} : $\alpha\ \mathtt{list}$ and returns a vector having each component of data-type $\mathds{\alpha}$. It utilizes the function $\mathtt{EL \ m \ L}$, which returns the $m^{th}$ element of a list $\texttt{L}$. Here, the \texttt{lambda} operator in HOL is used to construct a vector based on its components~\cite{harrison2013hol}.

\begin{mdframed}
\begin{definition}
\label{DEF:exp_ccos_csine}
\emph{Real Cosine and  Real Sine Functions} \\{
\textup{\texttt{$\vdash$ $\forall$ x. cos x = Re (ccos (Cx x)) \\
$\mathtt{}$$\vdash$ $\forall$ x. sin x = Re (csin (Cx x))
}}}
\end{definition}
\end{mdframed}

The real cosine and real sine are represented as $\texttt{cos}:\mathds{R} \rightarrow \mathds{R}$
and $\mathtt{sin}:\mathds{R} \rightarrow \mathds{R}$ in HOL Light~\cite{hol_light2017transcendentals}, respectively. These functions are formally defined using the complex cosine \texttt{ccos} : $\mathds{R}^2 \rightarrow \mathds{R}^2$ and complex sine \texttt{csin} : $\mathds{R}^2 \rightarrow \mathds{R}^2$ functions, respectively.

\begin{mdframed}
\begin{definition}
\label{DEF:vector_derivative}
\emph{Real Derivative} \\
{
\textup{\texttt{$\vdash$ $\forall$ f x. real\_derivative f x =
}}} \\
{\small
\textup{\texttt{ \hspace*{2.0cm} (@f$\mathtt{'}$. (f has\_real\_derivative f$\mathtt{'}$) (atreal x))
}}}
\end{definition}
\end{mdframed}

The function $\mathtt{real\_derivative}$ accepts a function $\texttt{f} : \mathds{R} \rightarrow \mathds{R}$ and a real number $\texttt{x}$, which is the point at which $\texttt{f}$ has to be differentiated, and returns a variable of data-type $\mathds{R}$, which represents the differential of $\texttt{f}$ at $\texttt{x}$. The function $\mathtt{has\_real\_derivative}$ defines the same relationship in the relational form.

We build upon the above-mentioned fundamental functions of multivariable calculus to formally analyze the robotic cell injection system in Sections~\ref{SEC:form_analy_cell_inj_sys} and~\ref{SEC:form_mot_pla_inj_pip} of the paper.


\subsection{Robotic Cell Injection Systems} \label{SUBSEC:rob_cell_inj_sys}

\begin{figure}[!ht]
\centering
\scalebox{0.85}
{\hspace*{-0.4cm} \includegraphics[trim={5.0 0.4cm 5.0 0.4cm},clip]{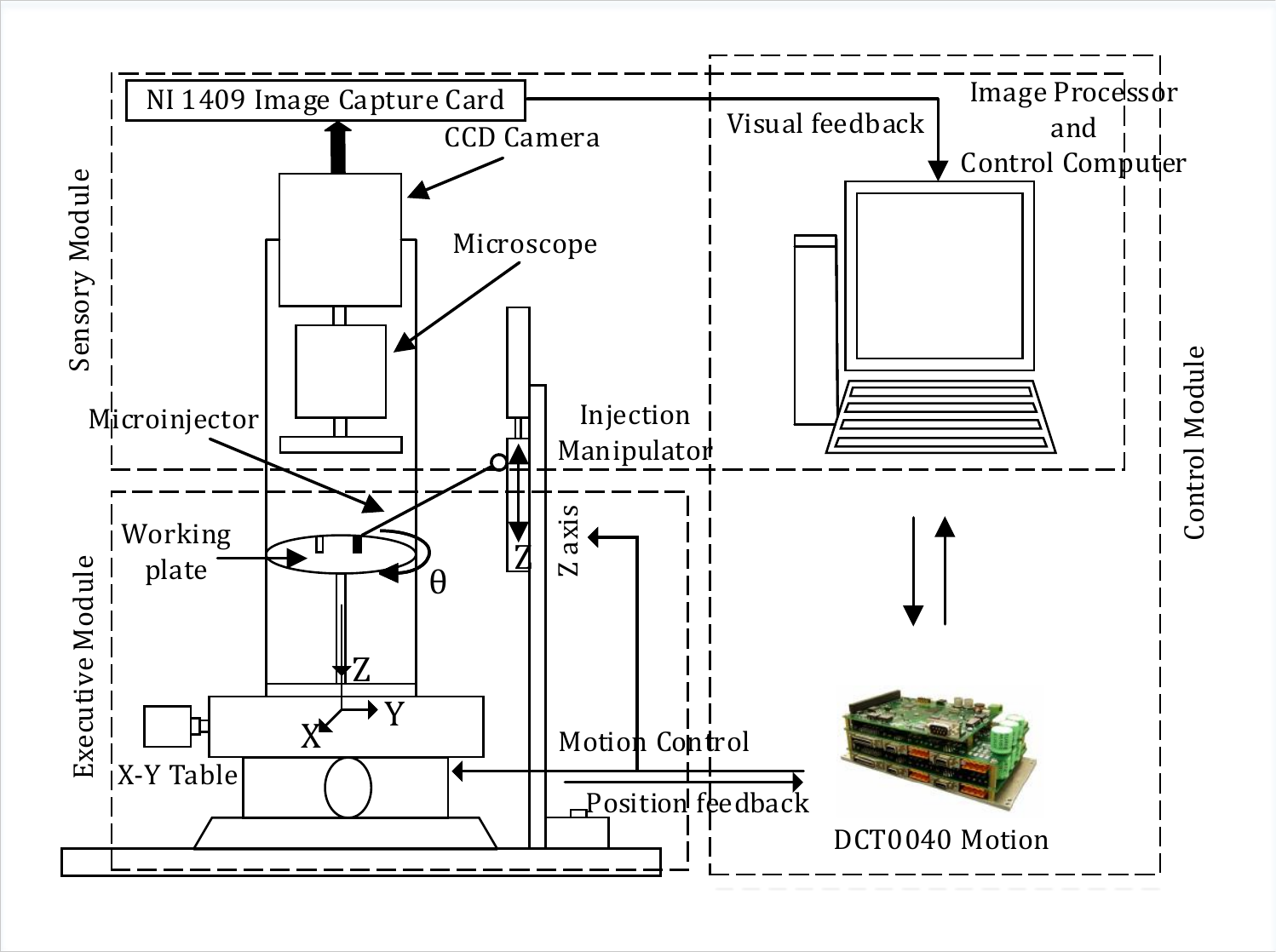}}
\caption{Robotic Cell Injection Systems}
\label{FIG:rob_cell_injec}
\end{figure}

A robotic cell injection system mainly comprises of three modules, namely executive, sensory and control modules as depicted in Figure~\ref{FIG:rob_cell_injec}. The executive module consists of positioning table, working plate and the injection manipulator. The cells that need to be injected are placed on a working plate, which is mounted on a positioning table ($XY\theta$-axis) and the injection manipulator is mounted on $Z$-axis as shown in Figure~\ref{FIG:rob_cell_injec}.

\begin{figure}[!ht]
\centering
\scalebox{0.80}
{\includegraphics[trim={5.0 0.4cm 5.0 0.4cm},clip]{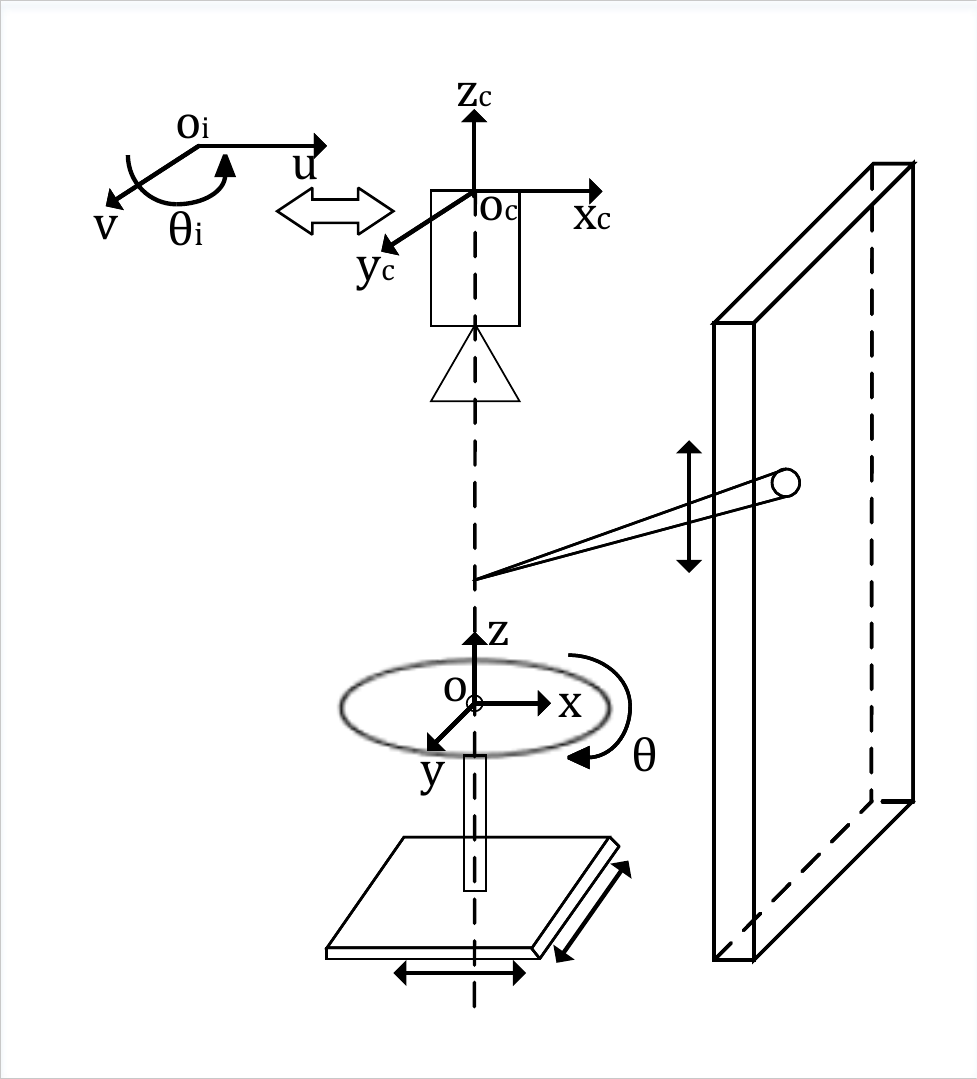}}
\caption{Configuration of the Robotic Cell Injection Systems}
\label{FIG:config_rob_cell_injec}
\end{figure}

The sensory module comprises of a vision system that has four parts, namely optical microscope, charged coupled device (CCD) camera, peripheral component interconnect (PCI) image capture and a processing card. The CCD camera is used to capture the cell injection process using a PCI image capture. The control module contains a host computer and a DCT0040 motion control system. Figure~\ref{FIG:config_rob_cell_injec} depicts the configuration of a robotic cell injection system. The axis $o-xyz$ represents the stage (table and working plate) coordinate frame, where $o$ is the origin of these coordinates representing the center of the working plate and $z$ is along the optical axis of the microscope. Similarly, $o_c-x_cy_cz_c$ is the camera coordinate frame with $o_c$ representing the center of the microscope. The coordinate frame in image plane is represented as $o_i-uv$, where $o_i$ is the origin and the axis $uv$ is perpendicular to the optical axis.

\section{Formalization of Robotic Cell Injection System}\label{SEC:form_analy_cell_inj_sys}

We present the higher-order-logic formalization of the robotic cell injection system using standard mathematical notations rather than the HOL Light notations, to facilitate the understanding of the paper for a non-HOL user.
The source code for our formalization can be obtained from~\cite{adnan17robcellinjsystp} for the readers who are interested to view the exact HOL Light formalization, presented in this paper.
We consider $2$-DOF to represent the dynamics of the robotic cell injection system.
The camera, stage and image coordinates are two-dimensional coordinates, which are modeled as follows in HOL Light:

\begin{mdframed}
\begin{flushleft}
\begin{definition}
\label{DEF:two_dim_coor}
\emph{Two-dimensional Coordinates} \\{  
\vspace*{0.2cm}
\textup{\texttt{$\vdash$ $\forall$ x y t. twod\_coord x y t = $\begin{bmatrix}\texttt{x(t)} \\ \texttt{y(t)}\end{bmatrix}$
}}}
\end{definition}
\end{flushleft}
\end{mdframed}

\noindent where \texttt{x} and \texttt{y} with data-type $\mathds{R} \rightarrow \mathds{R}$ representing the respective axes and \texttt{t} is a variable representing the time.

Next, we model the rotation matrix from the stage coordinate frame ($o-xyz$) to the camera coordinate frame ($o_c-x_xy_cz_c$), and the two-dimensional displacement vector between the origins of both these frames:

\begin{mdframed}
\begin{flushleft}
\begin{definition}
\label{DEF:rot_mat_n_disp_vec}
\emph{Rotation Matrix and Displacement Vector} \\{   
\vspace*{0.1cm}
\textup{\texttt{$\vdash$ $\forall$ alpha. rot\_mat alpha = 
$\begin{bmatrix}\texttt{cos (alpha)} & \hspace*{0.3cm} \texttt{sin (alpha)} \\ \texttt{-sin (alpha)} & \hspace*{0.3cm} \texttt{cos (alpha)} \end{bmatrix}$  \\ \vspace*{0.1cm}
$\mathtt{}$$\vdash$ $\forall$ dx dy.  disp\_vec dx dy =  $\begin{bmatrix}\texttt{dx} \\ \texttt{dy}\end{bmatrix}$
}}}
\end{definition}
\end{flushleft}
\end{mdframed}

The verification of the relationship between stage, camera and image coordinates provides key information for the reliable operation of the cell injection system by ensuring the accuracy of the orientation and movement of its various components, i.e., stage frame, microscope, camera and injection manipulator. Firstly, we verify the relationship between camera and stage coordinates as:

\hspace*{0.1cm}

\begin{mdframed}
\begin{flushleft}
\begin{theorem}
\label{THM:rel_camera_stage}
\emph{Relationship Between Camera and Stage Coordinates} \\{
\textup{\texttt{$\vdash$ $\forall$ xc yc x y alpha dx dy t. \\
$\mathtt{}$ \hspace*{0.2cm} \textbf{[A1]:} 0 < dx $\wedge$  \\ 
$\mathtt{}$ \hspace*{0.2cm} \textbf{[A2]:} 0 < dy   \\ \vspace*{0.1cm}
$\mathtt{}$ \hspace*{0.3cm} $\Rightarrow$ \Big(rel\_cam\_sta\_coord xc yc x y alpha dx dy t $\Leftrightarrow$   \\ \vspace*{0.1cm}
$\mathtt{}$ \hspace*{0.0cm}  $\begin{bmatrix}\texttt{xc(t)} \\ \texttt{yc(t)}\end{bmatrix}$ = $\begin{bmatrix}\texttt{x(t) $\ast$ cos (alpha) + y(t) $\ast$ sin (alpha) + dx} \\ \texttt{- x(t) $\ast$ sin (alpha) + y(t) $\ast$ cos (alpha) + dy}\end{bmatrix}$\Bigg)
}}}
\end{theorem}
\end{flushleft}
\end{mdframed}

\noindent where the HOL Light function \texttt{rel\_cam\_sta\_coord} models the relationship between camera and stage coordinates. The two assumptions of the above theorem provide the design constraints for the relationship.
The above theorem is verified using the properties of vectors and matrices alongside some real arithmetic reasoning.
Next, to verify the relationship between image and camera coordinates, we first model the display resolution matrix as the following HOL Light function:

\begin{mdframed}
\begin{flushleft}
\begin{definition}
\label{DEF:disp_res_mat}
\emph{Display Resolution Matrix} \\{    
\vspace*{0.1cm}
\textup{\texttt{$\vdash$ $\forall$ fx fy. disp\_res\_mat fx fy =
$\begin{bmatrix}\texttt{fx} & \hspace*{0.2cm} \texttt{0} \\ \texttt{0} & \hspace*{0.2cm} \texttt{fy} \end{bmatrix}$
}}}
\end{definition}
\end{flushleft}
\end{mdframed}

Now the image-camera coordinate frame interrelationship is verified as:

\begin{mdframed}
\begin{flushleft}
\begin{theorem}
\label{THM:rel_image_camera}
\emph{Relationship Between Image and Camera Coordinates} \\{
\textup{\texttt{$\vdash$ $\forall$ xc yc u v t fx fy.   \\
$\mathtt{}$ \hspace*{0.2cm} \textbf{[A1]:} 0 < fx $\wedge$  \\
$\mathtt{}$ \hspace*{0.2cm} \textbf{[A2]:} 0 < fy   \\ \vspace*{0.1cm}
$\mathtt{}$  \hspace*{0.1cm} $\Rightarrow$  \Bigg(rel\_ima\_cam\_coord xc yc u v t fx fy $\Leftrightarrow$   \\
\hspace*{5.5cm}  $\begin{bmatrix}\texttt{u(t)} \\ \texttt{v(t)}\end{bmatrix}$ = $\begin{bmatrix}\texttt{fx $\ast$ xc(t)} \\ \texttt{fy $\ast$ yc(t)}\end{bmatrix}$ \Bigg)
}}}
\end{theorem}
\end{flushleft}
\end{mdframed}

\noindent where the HOL Light function \texttt{rel\_ima\_cam\_coord} models the relationship between the image and the camera coordinates. The two assumptions of Theorem~\ref{THM:rel_image_camera} provide the design constraints for the relationship.
Next, we model the transformation matrix between image and stage coordinate frames, which is used in the verification of their interrelationship and is given as follows:

\begin{mdframed}
\begin{flushleft}
\begin{definition}
\label{DEF:trans_mat}
\emph{Transformation Matrix} \\{
\textup{\texttt{$\vdash$ $\forall$ fx fy alpha. transf\_mat fx fy alpha = \\ \vspace*{0.1cm}
\hspace*{3.5cm}
 $\begin{bmatrix}\texttt{fx $\ast$ cos (alpha)} & \hspace*{0.4cm} \texttt{fx $\ast$ sin (alpha)} \\ \texttt{-fy $\ast$ sin (alpha)} & \hspace*{0.4cm} \texttt{fy $\ast$ cos (alpha)} \end{bmatrix}$
}}}
\end{definition}
\end{flushleft}
\end{mdframed}

Now, we verify an important relationship between the image and stage coordinates as the following HOL Light theorem:

\begin{mdframed}
\begin{flushleft}
\begin{theorem}
\label{THM:rel_image_stage}
\emph{Relationship Between Image and Stage Coordinates} \\{
\textup{\texttt{$\vdash$ $\forall$ x y u v t fx fy dx dy alpha xc yc. \\
$\mathtt{}$ \hspace*{0.1cm} \textbf{[A1]:} 0 < dx $\wedge$  \\
$\mathtt{}$ \hspace*{0.1cm} \textbf{[A2]:} 0 < dy $\wedge$  \\
$\mathtt{}$ \hspace*{0.1cm} \textbf{[A3]:}  0 < fx $\wedge$  \\
$\mathtt{}$ \hspace*{0.1cm} \textbf{[A4]:} 0 < fy $\wedge$  \\
$\mathtt{}$ \hspace*{0.1cm} \textbf{[A5]:} twod\_coord u v t =  disp\_res\_mat fx fy $\ast\ast$  \\ \hspace*{6.0cm}  twod\_coord xc yc t $\wedge$  \\
$\mathtt{}$ \hspace*{0.1cm} \textbf{[A6]:} twod\_coord xc yc t = rot\_mat alpha $\ast\ast$  \\ \hspace*{4.0cm}  twod\_coord x y t + disp\_vec dx dy  \\
$\mathtt{}$ \hspace*{0.5cm} $\Rightarrow$   twod\_coord u v t = transf\_mat fx fy alpha $\ast\ast$ \\ \vspace*{0.1cm}
$\mathtt{}$ \hspace*{4.5cm} twod\_coord x y t + $\begin{bmatrix}\texttt{fx $\ast$ dx}  \\ \texttt{fy $\ast$ dy}  \end{bmatrix}$
}}}
\end{theorem}
\end{flushleft}
\end{mdframed}

\noindent where $\mathtt{\ast\ast}$ represents the matrix-vector multiplication. The first four assumptions (A1-A4) model the design constraints for the relationship between image and stage coordinates. The next assumption (A5) presents the relationship between image and camera coordinates. The last assumption (A6) presents the relationship between camera and stage coordinates. The verification of Theorem~\ref{THM:rel_image_stage} is mainly based on Theorems~\ref{THM:rel_camera_stage} and~\ref{THM:rel_image_camera}, and some classical properties of the vectors and matrices. The verification of these relationships raise our confidence about the orientation of the vital components of a cell injection system, i.e., injection manipulator, working plate, camera and microscope.

Next, we model and verify the dynamics of the cell injection system. The dynamics of the $2$-DOF motion stage, based on Lagrange's equation, is mathematically expressed as:

\begin{equation}\label{EQ:dyn_2_dof_stage_motion}
\begin{split}
\begin{bmatrix}m_x + m_y + m_p & \hspace*{0.3cm} 0  \\ 0 & \hspace*{0.3cm} m_y + m_p  \end{bmatrix} \begin{bmatrix} \dfrac{d^2x}{dt} \vspace*{0.3cm} \\ \dfrac{d^2y}{dt}  \end{bmatrix} + \begin{bmatrix}1 & \hspace*{0.2cm} 0  \\ 0 & \hspace*{0.2cm} 1  \end{bmatrix}  \begin{bmatrix} \dfrac{dx}{dt} \vspace*{0.3cm} \\ \dfrac{dy}{dt}  \end{bmatrix} =  \begin{bmatrix} \tau_x \vspace*{0.1cm}  \\ \tau_y  \end{bmatrix} - \begin{bmatrix} {fex}^d \vspace*{0.1cm} \\ {fey}^d  \end{bmatrix}
\end{split}
\end{equation}

\noindent where $m_x$, $m_y$ and $m_p$ are the masses of the $xy$ positioning tables and working plate, respectively. Similarly, $\tau_x$ and $\tau_y$ represent the components of the input torque to the driving motor. Similarly, ${fex}^d$ and ${fey}^d$ represent the components of the desired force applied to the actuators during the process of the cell injection. We formalize Equation~(\ref{EQ:dyn_2_dof_stage_motion}) as the following HOL Light function:

\begin{mdframed}
\begin{flushleft}
\begin{definition}
\label{DEF:dyn_2_dof}
\emph{Dynamics of the $2$-DOF Motion Stage} \\{   
\vspace*{0.1cm}
\textup{\texttt{$\vdash$ $\forall$ mx my mp x y t taux tauy fexd feyd. \\
\hspace*{0.3cm} dyn\_2\_dof\_mot\_sta mx my mp x y t taux tauy fexd feyd $\Leftrightarrow$ \\
\hspace*{0.5cm} mass\_mat mx my mp $\ast\ast$ sec\_ord\_der\_sta\_coord x y t + \\
\hspace*{2.0cm} pos\_tab\_mat $\ast\ast$ fir\_ord\_der\_sta\_coord x y t =  \\
\hspace*{3.0cm} tor\_vec taux tauy - des\_force\_vec fexd feyd
}}}
\end{definition}
\end{flushleft}
\end{mdframed}

\noindent where \texttt{mass\_mat} is the matrix containing the respective masses and \texttt{pos\_tab\_mat} is the diagonal matrix. Similarly, \texttt{tor\_vec} and \texttt{des\_force\_vec} are the vectors with their elements representing the components of the applied torque and desired force. The HOL Light functions \texttt{fir\_ord\_der\_sta\_coord} and \texttt{sec\_ord\_der\_s} \texttt{ta\_coord} model the vectors having first-order and second-order derivatives of the stage coordinates:

\begin{mdframed}
\begin{flushleft}
\begin{definition}
\label{DEF:fst_n_scd_ord_der_vec}
\emph{First and Second-order Derivative Vectors} \\{   
\vspace*{0.1cm}
\textup{\texttt{$\vdash$ $\forall$ x y t. fir\_ord\_der\_sta\_coord x y t = deriv\_vec\_fir [x; y] t \\
$\vdash$ $\forall$ x y t. sec\_ord\_der\_sta\_coord x y t = deriv\_vec\_sec [x; y] t
}}}
\end{definition}
\end{flushleft}
\end{mdframed}

\noindent where \texttt{deriv\_vec\_fir} and \texttt{deriv\_vec\_sec} accept a list containing the functions of data-type $\mathds{R} \rightarrow \mathds{R}$ and return the corresponding first and second-order derivative vectors~\cite{adnan17robcellinjsystp}.

If the applied torque and force vectors are zero, then the injection pipette does not touch the cells. Thus, Equation~(\ref{EQ:dyn_2_dof_stage_motion}) can be transformed for this particular scenario as follows:

\begin{equation}\label{EQ:dyn_2_dof_stage_motion_homog}
\begin{split}
\begin{bmatrix}m_x + m_y + m_p &  \hspace*{0.3cm} 0  \\ 0 & \hspace*{0.3cm} m_y + m_p  \end{bmatrix} \begin{bmatrix} \dfrac{d^2x}{dt} \vspace*{0.3cm} \\ \dfrac{d^2y}{dt}  \end{bmatrix} + \begin{bmatrix}1 & \hspace*{0.2cm} 0  \\ 0 & \hspace*{0.2cm} 1  \end{bmatrix}
\begin{bmatrix} \dfrac{dx}{dt} \vspace*{0.3cm} \\ \dfrac{dy}{dt}  \end{bmatrix} = \begin{bmatrix} 0  \\ 0  \end{bmatrix} \end{split}
\end{equation}

We verify the solution of the above equation as the following HOL Light theorem:

\begin{mdframed}
\begin{flushleft}
\begin{theorem}
\label{THM:soln_ver_dyn_stage_motion_homog}
\emph{Verification of Solution of Dynamical Behaviour of Motion Stage} \\{
\textup{\texttt{$\vdash$ $\forall$ x y mx my mp taux tauy fexd feyd alpha x0 y0 xd0 yd0. \\
$\mathtt{}$ \hspace*{0.1cm} \textbf{[A1]:} 0 < mx  $\wedge$ \textbf{[A2]:} 0 < my  $\wedge$ \textbf{[A3]:} 0 < mp $\wedge$  \\
$\mathtt{}$ \hspace*{0.1cm} \textbf{[A4]:} x(0) = x0 $\wedge$ \textbf{[A5]:} y(0) = y0 $\wedge$  \\ \vspace*{0.1cm}
$\mathtt{}$ \hspace*{0.1cm} \textbf{[A6]:} $\mathtt{\dfrac{dx}{dt}}$(0)= xd0 $\wedge$ \textbf{[A7]:} $\mathtt{\dfrac{dy}{dt}}$(0)= yd0 $\wedge$  \\ \vspace*{0.2cm}
$\mathtt{}$ \hspace*{0.1cm}  \textbf{[A8]:} $\mathtt{}$   $\begin{bmatrix}\texttt{taux}  \\ \texttt{tauy}  \end{bmatrix}$ = $\begin{bmatrix}\texttt{0}  \\ \texttt{0}  \end{bmatrix}$ $\wedge$  \\ \vspace*{0.1cm}
$\mathtt{}$ \hspace*{0.1cm}  \textbf{[A9]:} $\mathtt{}$   $\begin{bmatrix}\texttt{fexd}  \\ \texttt{feyd}  \end{bmatrix}$ = $\begin{bmatrix}\texttt{0}  \\ \texttt{0}  \end{bmatrix}$ $\wedge$  \\ \vspace*{0.1cm}
$\mathtt{}$ \hspace*{0.1cm}  \textbf{[A10]:} ($\forall$ t. x(t) = (x0 + xd0 $\ast$ (mx + my + mp))  \\  \vspace*{0.1cm}
$\mathtt{}$  \hspace*{4.0cm}  - xd0 $\ast$ (mx + my + mp) $\ast$  \large$\mathtt{e^{{\frac{-1}{mx + my + mp}} t}}$ $\wedge$  \\  \vspace*{0.1cm}
$\mathtt{}$ \hspace*{0.05cm}  \textbf{[A11]:} ($\forall$ t. y(t) = (y0 + yd0 $\ast$ (my + mp))  \\ \vspace*{0.2cm}
$\mathtt{}$  \hspace*{4.0cm}  - yd0 $\ast$ (my + mp) $\ast$  \large{$\mathtt{e^{{\frac{-1}{my + mp}} t}}$}   \\ \vspace*{0.1cm}
$\mathtt{}$  \hspace*{1.0cm} $\Rightarrow$ dyn\_2\_dof\_mot\_sta mx my mp x y t taux tauy fexd feyd
}}}
\end{theorem}
\end{flushleft}
\end{mdframed}

\noindent  The first three assumptions (A1-A3) model the condition that all the masses, i.e., \texttt{mx}, \texttt{my} and \texttt{mp} are positive. The next four assumptions (A4-A7) present the values of coordinates \texttt{x} and \texttt{y} and their first-order derivatives $\mathtt{\frac{dx}{dt}}$ and $\mathtt{\frac{dy}{dt}}$ at $t = 0$. The next two assumptions (A8-A9) model the condition that the torque and force vectors are zero. The next two assumptions (A10-A11) provide the values of $xy$ coordinates at any time $t$. Finally, the conclusion presents the dynamics of the $2$-DOF motion stage. The proof-process of Theorem~\ref{THM:soln_ver_dyn_stage_motion_homog} involves the properties of real derivatives, transcendental functions, matrices and vectors alongwith some real arithmetic reasoning. Next, we verify an alternate form of the relationship between the image and stage coordinates, which depends on the dynamics of the motion stage (Definition~\ref{DEF:dyn_2_dof}) and is a vital property for the analysis of cell injection systems. For this purpose, we first model the positioning table matrix and inertia matrix:

\begin{mdframed}
\begin{flushleft}
\begin{definition}
\label{DEF:pos_tab_n_iner_mat}
\emph{Positioning Table and Inertia Matrices} \\{
\textup{\texttt{$\vdash$ $\forall$ fx fy alpha.   \\
$\mathtt{}$ \hspace*{0.6cm} pos\_tab\_mat\_fin fx fy alpha =  \\
$\mathtt{}$  \hspace*{2.0cm} pos\_tab\_mat $\ast\ast$ matrix\_inv (transf\_mat fx fy alpha) \\
$\mathtt{}$$\vdash$ $\forall$ mx my mp fx fy alpha.  \\
$\mathtt{}$ \hspace*{0.6cm}  iner\_mat mx my mp fx fy alpha = \\
$\mathtt{}$ \hspace*{0.8cm} mass\_mat mx my mp $\ast\ast$ matrix\_inv (transf\_mat fx fy alpha)
}}}
\end{definition}
\end{flushleft}
\end{mdframed}

\noindent where the HOL Light function \texttt{matrix\_inv} accepts a matrix \texttt{A:}${\mathds{R}^M}^N$ and returns its inverse. Now, the alternate representation of the image-stage coordinate frame interrelationship is verified as the following HOL Light theorem:

\begin{mdframed}
\begin{flushleft}
\begin{theorem}
\label{THM:rel_image_stage_alt}
\emph{Alternate Form of Relationship Between Image and Stage Coordinates} \\{
\textup{\texttt{$\vdash$ $\forall$ xc yc u v x y fx fy dx dy mx my mp taux tauy fexd feyd alpha. \\
$\mathtt{}$ \hspace*{0.1cm} \textbf{[A1]:} 0 < dx  $\wedge$ \textbf{[A2]:} 0 < dy  $\wedge$  \\
$\mathtt{}$ \hspace*{0.1cm} \textbf{[A3]:} 0 < fx  $\wedge$ \textbf{[A4]:} 0 < fy  $\wedge$  \\
$\mathtt{}$ \hspace*{0.27cm}\textbf{[A5]:} invertible (transf\_mat fx fy alpha) $\wedge$  \\
$\mathtt{}$ \hspace*{0.27cm}\textbf{[A6]:} ($\forall$ t. u real\_differentiable atreal t) $\wedge$   \\
$\mathtt{}$ \hspace*{0.27cm}\textbf{[A7]:} ($\forall$ t. v real\_differentiable atreal t) $\wedge$   \\ \vspace*{0.1cm}
$\mathtt{}$ \hspace*{0.27cm}\textbf{[A8]:} ($\forall$ t. $\mathtt{\dfrac{du}{dt}}$ real\_differentiable atreal t) $\wedge$   \\ \vspace*{0.1cm}
$\mathtt{}$ \hspace*{0.27cm}\textbf{[A9]:} ($\forall$ t. $\mathtt{\dfrac{dv}{dt}}$ real\_differentiable atreal t) $\wedge$   \\ \vspace*{0.1cm}
$\mathtt{}$ \hspace*{0.27cm}\textbf{[A10]:} ($\forall$ t. rel\_ima\_cam\_coord xc yc u v t fx fy) $\wedge$  \\
$\mathtt{}$ \hspace*{0.27cm}\textbf{[A11]:} ($\forall$ t. rel\_cam\_sta\_coord xc yc x y alpha dx dy t) $\wedge$  \\
$\mathtt{}$ \hspace*{0.27cm}\textbf{[A12]:} dyn\_2\_dof\_mot\_sta mx my mp x y t taux tauy fexd feyd   \\
$\mathtt{}$   \hspace*{1.0cm}  $\Rightarrow$ iner\_mat mx my mp fx fy alpha $\ast\ast$   \\
$\mathtt{}$  \hspace*{4.5cm} sec\_ord\_der\_ima\_coord u v t +  \\
$\mathtt{}$   \hspace*{1.5cm}   pos\_tab\_mat\_fin fx fy alpha $\ast\ast$  \\
$\mathtt{}$   \hspace*{4.5cm}   fir\_ord\_der\_ima\_coord u v t =  \\
$\mathtt{}$   \hspace*{3.0cm}   tor\_vec taux tauy - des\_force\_vec fexd feyd
}}}
\end{theorem}
\end{flushleft}
\end{mdframed}

The first four assumptions (A1-A4) describe the design constraints for the image-stage interrelationship. The next assumption (A5) ensures that the transformation matrix (\texttt{transf\_mat}, Definition~\ref{DEF:trans_mat}) is invertible, i.e., its inverse exists. The next four assumptions (A6-A9) model the differentiability condition for the image coordinates and their first-order derivatives.
The next two assumptions (A10-A11) provide the image-camera and camera-stage coordinate frames interrelationships. The last assumption (A12) represents the dynamics of the $2$-DOF motion stage. Finally, the conclusion of Theorem~\ref{THM:rel_image_stage_alt} is the alternate representation of the image-stage coordinate frame interrelationship. The verification of Theorem~\ref{THM:rel_image_stage_alt} is based on the properties of the real derivative, matrices and vectors alongwith some real arithmetic reasoning.

\section{Formalization of the Motion Planning of the Injection Pipette}\label{SEC:form_mot_pla_inj_pip}

The injection motion controller is another vital part of the cell injection systems and its verification is necessary for a reliable system. It mainly includes the control of the applied injection force and the torque applied to the deriving motor. So, we formalize the force and torque controls and formally verify the implication relationship between both of these controllers. The impendence force control for a cell injection system is represented as follows:

\begin{equation}\label{EQ:imped_force_control}
\begin{split}
m\ddot{e} + b\dot{e} + ke = f_e
\end{split}
\end{equation}

\noindent where $m$, $b$ and $k$ represent the desired impendence parameters. Similarly, $f_e$ is the two-dimensional vector having $f_{ex}$ and $f_{ey}$ as its elements, which represent the $x$ and $y$ components of the applied force. Moreover, $e$, $\dot{e}$ and $\ddot{e}$ are the vectors representing the position errors of the $xy$ stage coordinates, their first-order and second-order derivatives, respectively, and are mathematically expressed as:

\vspace*{0.1cm}

\begin{equation}\label{EQ:pos_error_vec}
\begin{split}
e = \begin{bmatrix}  x_d \\ y_d   \end{bmatrix} - \begin{bmatrix}  x \\ y   \end{bmatrix}, \
\dot{e} = \begin{bmatrix} \dfrac{d{x_d}}{dt} \vspace*{0.3cm} \\ \dfrac{d{y_d}}{dt}  \end{bmatrix} - \begin{bmatrix} \dfrac{dx}{dt} \vspace*{0.3cm} \\ \dfrac{dy}{dt}  \end{bmatrix}, \
\ddot{e} = \begin{bmatrix} \dfrac{d^2{x_d}}{dt} \vspace*{0.3cm} \\ \dfrac{d^2y_d}{dt}  \end{bmatrix} - \begin{bmatrix} \dfrac{d^2x}{dt} \vspace*{0.3cm} \\ \dfrac{d^2y}{dt}  \end{bmatrix}
\end{split}
\end{equation}

\noindent where $x$ and $y$ are the actual axes and $x_d$ and $y_d$ are the desired axes of the stage coordinate frame. Now, the image-based torque controller for the $xy$ stage coordinates is mathematically expressed as:

\vspace*{0.1cm}

\begin{equation}\label{EQ:torque_controller}
\begin{split}
\begin{bmatrix} \tau_x \vspace*{0.1cm} \\ \tau_y  \end{bmatrix} =
\begin{bmatrix}m_x + m_y + m_p \vspace*{0.1cm} & \hspace*{0.3cm} 0  \\ 0 \vspace*{0.1cm} & \hspace*{0.3cm} m_y + m_p  \end{bmatrix}
\begin{bmatrix}f_x \cos \alpha \vspace*{0.1cm} & \hspace*{0.3cm} f_x \sin \alpha  \\ -f_y \sin \alpha \vspace*{0.1cm} & \hspace*{0.3cm} f_y \cos \alpha  \end{bmatrix}
\begin{bmatrix} \dfrac{d^2x_d}{dt} \vspace*{0.3cm} \\ \dfrac{d^2y_d}{dt} \end{bmatrix} + \\ \vspace*{0.6cm}
\begin{bmatrix}m_x + m_y + m_p \vspace*{0.1cm} & \hspace*{0.3cm} 0  \\ 0 \vspace*{0.1cm} & \hspace*{0.3cm} m_y + m_p  \end{bmatrix}
\begin{bmatrix}f_x \cos \alpha \vspace*{0.1cm} & \hspace*{0.3cm} f_x \sin \alpha  \\ -f_y \sin \alpha \vspace*{0.1cm} & \hspace*{0.3cm} f_y \cos \alpha  \end{bmatrix}  \\ \vspace*{0.6cm}
m^{-1}(b\dot{e} + ke - f_e) + \Bigg( \begin{bmatrix} 1 \vspace*{0.1cm} & \hspace*{0.3cm} 0  \\ 0 \vspace*{0.1cm} & \hspace*{0.3cm} 1  \end{bmatrix}
{\begin{bmatrix}f_x \cos \alpha \vspace*{0.1cm} & \hspace*{0.3cm} f_x \sin \alpha  \\ -f_y \sin \alpha \vspace*{0.1cm} & \hspace*{0.3cm} f_y \cos \alpha  \end{bmatrix} }^{-1}\Bigg) \\ \vspace*{1.7cm}
\begin{bmatrix}f_x \cos \alpha \vspace*{0.1cm} & \hspace*{0.3cm} f_x \sin \alpha  \\ -f_y \sin \alpha \vspace*{0.1cm} & \hspace*{0.3cm} f_y \cos \alpha  \end{bmatrix}
\begin{bmatrix} \dfrac{dx}{dt} \vspace*{0.3cm} \\ \dfrac{dy}{dt}  \end{bmatrix} +
\begin{bmatrix} {fex}^d \vspace*{0.1cm} \\ {fey}^d  \end{bmatrix}
\end{split}
\end{equation}

\vspace*{0.1cm}

Equation~(\ref{EQ:torque_controller}) can be alternatively written as:

\vspace*{0.2cm}

\begin{equation}\label{EQ:torq_contr_vec_mat_form}
\begin{split}
\overrightarrow{\tau} = M T
\begin{bmatrix} \dfrac{d^2x_d}{dt} \vspace*{0.3cm}  \\ \dfrac{d^2y_d}{dt} \end{bmatrix} + M T
m^{-1}(b\dot{e} + ke - f_e) +
N T
\begin{bmatrix} \dfrac{dx}{dt}  \vspace*{0.3cm} \\ \dfrac{dy}{dt}  \end{bmatrix} +
\overrightarrow{f_{ed}}
\end{split}
\end{equation}

\hspace*{0.1cm}

\noindent where $M$, $N$ and $T$ in the above equation denote the inertia, positioning table and transformation matrices. The above equation was wrongly presented in simulations~\cite{huang2006visual} and model checking~\cite{sardar2017towards} based analysis as follows:

\vspace*{0.2cm}

\begin{equation}\label{EQ:torq_contr_vec_trad_rep}
\begin{split}
\overrightarrow{\tau} = M
\begin{bmatrix} \dfrac{d^2x_d}{dt}  \vspace*{0.3cm} \\ \dfrac{d^2y_d}{dt} \end{bmatrix} + M
m^{-1}(b\dot{e} + ke - f_e) +
N
\begin{bmatrix} \dfrac{dx}{dt}  \vspace*{0.3cm} \\ \dfrac{dy}{dt}  \end{bmatrix} +
\overrightarrow{f_{ed}}
\end{split}
\end{equation}

\vspace*{0.2cm}

\begin{equation}\label{EQ:torq_contr_vec_mc_rep}
\begin{split}
\overrightarrow{\tau} = M T
\begin{bmatrix} \dfrac{d^2x_d}{dt}  \vspace*{0.3cm} \\ \dfrac{d^2y_d}{dt} \end{bmatrix} + M T
m^{-1}(b\dot{e} + ke - f_e) +
N T
\begin{bmatrix} \dfrac{dx}{dt} \vspace*{0.3cm} \\ \dfrac{dy}{dt}  \end{bmatrix} +
\overrightarrow{f_{e}}
\end{split}
\end{equation}

\vspace*{0.2cm}

In Equation~(\ref{EQ:torq_contr_vec_trad_rep}) (used in the simulations based analysis~\cite{huang2006visual}), the transformation matrix (T) is missing, which includes the amount of applied force and the angles at which the injection pipette is pierced into the cell and its absence can lead to disastrous consequences, i.e., excess substance injection, damaging cell tissues etc. Similarly, in Equation~(\ref{EQ:torq_contr_vec_mc_rep}) (used in the model checking based analysis~\cite{sardar2017towards}), $f_{ed}$ is wrongly interpreted as $f_e$, i.e., the desired force, is taken equal to the applied force, which can never happen in a real-world system.
We caught these wrong interpretations of Equation~(\ref{EQ:torq_contr_vec_mat_form}) in the simulations and model checking based analyses during the verification of the implication relationship between force control and torque controller. We first started the verification of this relationship using Equation~(\ref{EQ:torq_contr_vec_trad_rep}) and ended up with the identification of this issue. Next, we took Equation~(\ref{EQ:torq_contr_vec_mc_rep}) and again, during its verification, identified its wrong interpretation, which enabled us to obtain its right interpretation as given in Equation~(\ref{EQ:torq_contr_vec_mat_form}).
We verified the image-based torque controller (Equation~(\ref{EQ:torq_contr_vec_mat_form})) as the following HOL Light theorem:

\vspace*{0.5cm}

\begin{mdframed}
\begin{flushleft}
\begin{theorem}
\label{THM:ver_toque_contr}
\emph{Verification of the Implication Relationship Between Force Control and Torque Controller} \\{
\textup{\texttt{$\vdash$ $\forall$ xd yd x y t mx my mp fx fy  \\
$\mathtt{}$ \hspace*{1.5cm} alpha taux tauy fex fey fexd feyd m b k. \\
$\mathtt{}$ \hspace*{0.1cm} \textbf{[A1]:} 0 < m  $\wedge$  \\
$\mathtt{}$ \hspace*{0.1cm} \textbf{[A2]:} 0 < k  $\wedge$  \\
$\mathtt{}$ \hspace*{0.1cm} \textbf{[A3]:} 0 < b $\wedge$  \\
$\mathtt{}$ \hspace*{0.1cm} \textbf{[A4]:} invertible (transf\_mat fx fy alpha) $\wedge$  \\
$\mathtt{}$ \hspace*{0.1cm} \textbf{[A5]:} force\_cont xd yd x y t m b k fex fey $\wedge$   \\
$\mathtt{}$ \hspace*{0.1cm} \textbf{[A6]:} dyn\_2\_dof\_mot\_sta mx my mp x y t taux tauy fexd feyd   \\
$\mathtt{}$  \hspace*{0.5cm} $\Rightarrow$ torque\_cont xd yd x y t mx my mp fx fy  \\
$\mathtt{}$  \hspace*{3.0cm} alpha taux tauy fex fey fexd feyd m b k
}}}
\end{theorem}
\end{flushleft}
\end{mdframed}

\vspace*{0.2cm}

The first three assumptions (A1-A3) ensure that the desired impendence parameters are positive. The next assumption (A4) provides the condition that the transformation matrix (\texttt{transf\_mat}) is invertible. The next assumption (A5) models the impendence force control (Equation~(\ref{EQ:imped_force_control})). The last assumption (A6) presents the dynamics of the $2$-DOF motion stage. Finally, the conclusion represents the image-based torque controller (Equation~(\ref{EQ:torque_controller})). The verification of Theorem~\ref{THM:ver_toque_contr} is mainly based on the properties of real derivative, vector and matrices.

Due to the undecidable nature of the higher-order logic, the verification results presented in Sections~\ref{SEC:form_analy_cell_inj_sys} and~\ref{SEC:form_mot_pla_inj_pip}, involved manual interventions and human guidance. However, we developed some tactics to automate the verification process. For example, we developed a tactic \texttt{VEC\_MAT\_SIMP\_TAC}, which simplifies the matrix and vector arithmetics involved in the formal analysis of the robotic cell injection system. Thus, the proof effort involved only $745$ lines-of-code and $17$ man-hours. The details about these tactics and rest of the formalization can be found in our proof script~\cite{adnan17robcellinjsystp}. The distinguishing feature of our formal analysis is that all the verified theorems are universally quantified and can thus be specialized to the required values based on the requirement of the analysis of the cell injection systems. Moreover, our approach allows us to model the dynamics of the cell injection systems involving differential and derivative (Equations~(\ref{EQ:dyn_2_dof_stage_motion}),~(\ref{EQ:imped_force_control}),~(\ref{EQ:torque_controller})) in their true form, whereas, in their model checking based analysis~\cite{sardar2017towards}, they are discretized and modeled using a state-transition system, which may compromise the accuracy and completeness of the corresponding analysis.

\section{Conclusion}\label{SEC:conclusion}

In this paper, we presented a formal analysis of robotic cell injection systems. We first formalize the stage, camera and image coordinate frames, which are the main components of a robotic cell injection system, and formally verified their interrelationship using the HOL Light theorem prover. We also formalized the dynamics of the $2$-DOF motion stage based on differential equations and verified their solutions in HOL Light. Finally, we formalized the impedance force control and image-based torque controller and verified their implication relationship. Our formalization helped us to identify some key discrepancies in the simulation-based and model checking based analysis of these systems, which shows the usefulness of using higher-order-logic theorem proving in the formal analysis of critical systems.

\bibliographystyle{splncs03}
\bibliography{biblio}

\end{document}